\documentclass{PoS}
\usepackage{cite}

\title{Theoretical results for electroweak-boson and single-top production}

\ShortTitle{Theoretical results for electroweak-boson and single-top production}

\author{\speaker{Nikolaos Kidonakis}\thanks{This material is based upon work supported by the National Science Foundation under Grant No. PHY 1212472.}\\
    Department of Physics, Kennesaw State University, Kennesaw, GA 30144, USA\\
        E-mail: \email{nkidonak@kennesaw.edu}}

\abstract{I present results from recent high-order calculations for the production of electroweak bosons and top quarks. In particular, I discuss $W$ and $Z$ boson production at large transverse momentum, single-top production, and FCNC top production. Theoretical predictions which include higher-order soft-gluon corrections are presented for total cross sections and differential distributions at the LHC.}

\FullConference{XXIII International Workshop on Deep Inelastic Scattering\\
		27 April - May 1 2015\\
		Dallas, Texas}

\def\beq{\begin{equation}}
\def\eeq{\end{equation}}
\def\beqa{\begin{eqnarray}}
\def\eeqa{\end{eqnarray}}

\begin{document}

\section{Introduction}

Higher-order corrections are very significant for $W$ and $Z$ distributions at large transverse momentum, $p_T$, as well as for single-top production cross sections and $p_ T$ distributions. Soft-gluon contributions are an important and dominant part of the perturbative corrections.

The soft-gluon terms in the $n$th-order corrections involve logarithms 
$[\ln^k(s_4/p_T^2)/s_4]_+$  for electroweak-boson production and
$[\ln^k(s_4/m_t^2)/s_4]_+$  for single-top production,
with $k \le 2n-1$ and $s_4$ the kinematical distance from partonic threshold.
These soft corrections have been resummed at NNLL accuracy via the calculation of the corresponding two-loop soft anomalous dimensions.
Approximate NNLO (aNNLO) differential cross sections have been derived from the expansion of the resummed expressions for electroweak-boson \cite{NKRG,NKRGw}
and single-top \cite{NKtch,NKsch,NKtW,NKppn} production.

\section{$W$ and $Z$  production at large $p_T$}

The production of $W$ and $Z$ bosons is useful in testing the Standard Model and in estimates of backgrounds to Higgs production and new physics.  
The partonic channels at leading order are 
$q(p_a) + g(p_b) \longrightarrow W(Q) + q(p_c)$ and 
$q(p_a) + {\bar q}(p_b) \longrightarrow W(Q) + g(p_c)$. 
We define $s=(p_a+p_b)^2$, $t=(p_a-Q)^2$, $u=(p_b-Q)^2$
and $s_4=s+t+u-Q^2$. At threshold $s_4 \rightarrow 0$ 
and the soft corrections are of the form 
$[\ln^k(s_4/p_T^2)/s_4]_+$.
The latest aNNLO results at NNLL accuracy have been derived in \cite{NKRG}.

\begin{figure}
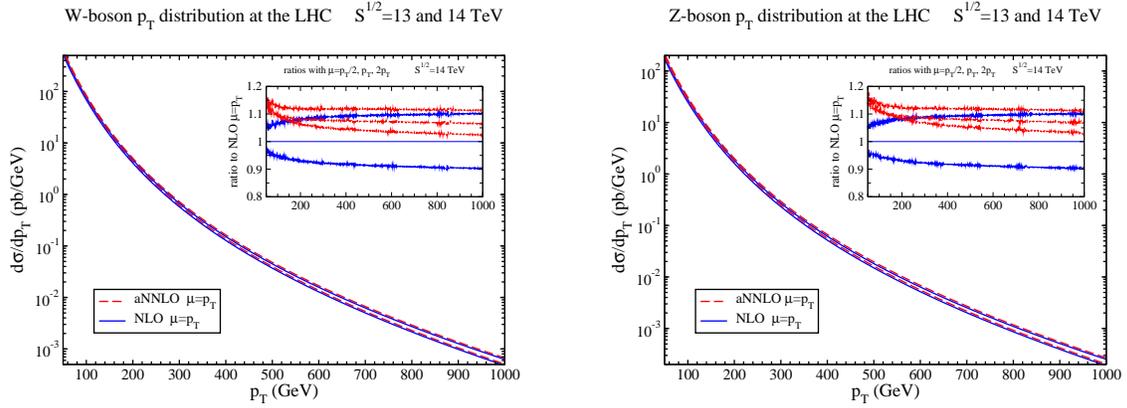

\begin{center}
\includegraphics[width=.45\textwidth]{W13and14lhcplot.eps}
\hspace{10mm}
\includegraphics[width=.45\textwidth]{Z13and14lhcplot.eps}
\caption{aNNLO $W$-boson (left) and $Z$-boson (right) $p_T$ distributions at the LHC.}
\label{figWZlhc}
\end{center}
\end{figure}

We begin with $W$ production at large $p_T$. We use MSTW2008 pdf \cite{MSTW2008} in our calculations. 
In the left plot of Fig. \ref{figWZlhc} we show the aNNLO $W$-boson $p_T$ distributions at the LHC at 13 and 14 TeV energies.  The $p_T$ 
distributions fall rapidly as the $p_T$ of the $W$ boson increases.
The inset plot displays the ratio of the results to the central NLO result. 
We observe the significant contribution of the aNNLO corrections and the reduction of scale uncertainty at aNNLO.

We continue with $Z$ production at large $p_T$. 
In the right plot of Fig. \ref{figWZlhc} we show the aNNLO $Z$-boson $p_T$ distributions at LHC energies of 13 and 14 TeV. Our observations regarding the size of the corrections and the reduction of uncertainty at aNNLO are the same as for $W$ production.

\section{Single-top production}

Single-top production can proceed via three partonic channels. At lowest order, the $t$-channel partonic processes are $qb \rightarrow q' t$ and ${\bar q} b \rightarrow {\bar q}' t$;
the $s$-channel processes are $q{\bar q}' \rightarrow {\bar b} t$;
and the associated $tW$ production processes are $bg \rightarrow tW^-$, and similarly for single antitop. NNLL resummation for all these processes was performed in Refs. \cite{NKtch,NKsch,NKtW,NKppn}.

\begin{table}[htb]
\begin{center}
\begin{tabular}{c|c|c|c}
LHC  & $t$ &  
${\bar t}$ & Total (pb) \\ 
\hline
8 TeV  & $55.9 {}^{+2.1}_{-0.3} \pm 1.1$
& $30.6 \pm 0.7 {}^{+0.9}_{-1.1}$
& $86.5 {}^{+2.8}_{-1.0} {}^{+2.0}_{-2.2}$
\\ 
13 TeV & $136 {}^{+3}_{-1} \pm 3$ 
& $82 {}^{+2}_{-1} \pm 2$ 
&$218 {}^{+5}_{-2} \pm 5$
\\ 
14 TeV & $154 {}^{+4}_{-1} \pm 3$ 
& $94 {}^{+2}_{-1} {}^{+2}_{-3}$ 
& $248 {}^{+6}_{-2} {}^{+5}_{-6}$
\end{tabular}
\caption{aNNLO single-top and single-antitop $t$-channel cross sections with $m_t=173.3$ GeV.}
\label{tab1}
\end{center}
\end{table}

In Table \ref{tab1} we show the single-top and single-antitop $t$-channel production cross sections at aNNLO at the LHC, as well as the total sum of the two.
The first uncertainty is from scale variation over $m_t/2 \le \mu \le 2m_t$ 
while the second is from 
pdf errors with MSTW2008 NNLO pdf at 90\% C.L. \cite{MSTW2008}. 

\begin{figure}
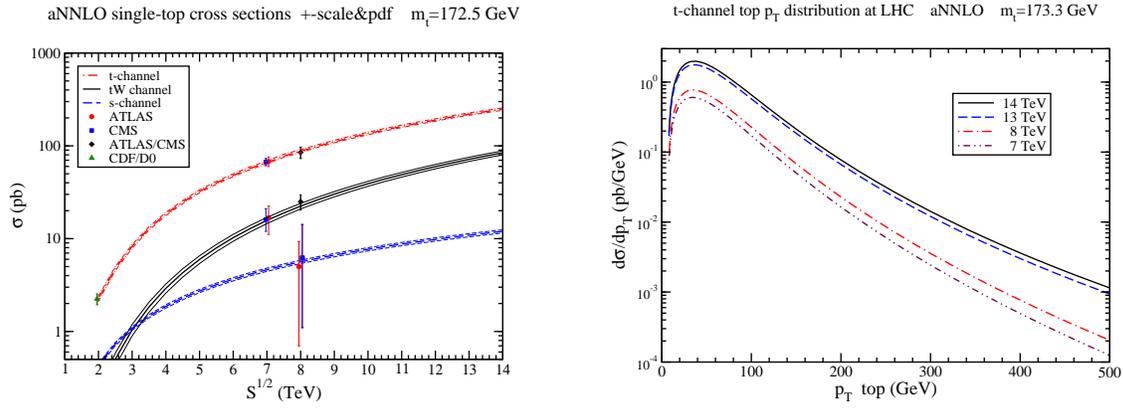

\begin{center}
\includegraphics[width=.45\textwidth]{singletopplot.eps}
\hspace{10mm}
\includegraphics[width=.45\textwidth]{pttoptchlhcplot.eps}
\caption{(Left) Single-top aNNLO cross sections compared with LHC \cite{ATLAStch7lhc,CMStch7lhc,tch8lhccombo,ATLASsch8lhc,CMSsch8lhc,ATLAStW7lhc,CMStW7lhc,tW8lhccombo} and Tevatron \cite{tchtevcombo} data; 
(Right) aNNLO $t$-channel top $p_T$ distributions at LHC energies.}
\label{figsingletop}
\end{center}
\end{figure}

In the left plot of Fig. \ref{figsingletop} we plot the $t$-channel total cross section as a function of collider energy.
At 7 TeV LHC energy, we compare with $t$-channel data from ATLAS \cite{ATLAStch7lhc} and CMS \cite{CMStch7lhc}.
At 8 TeV LHC energy we compare with ATLAS/CMS combination $t$-channel data \cite{tch8lhccombo}. Finally, at 1.96 Tevatron energy we compare with CDF/D0 combination data \cite{tchtevcombo}.
We find excellent agreement of theory with data for all collider energies.

The aNNLO ratio $\sigma(t)/\sigma({\bar t})= 1.82{}^{+0.10}_{-0.09}$ at 8 TeV 
which compares very well with the CMS result  $1.95 \pm 0.10 \pm 0.19$ \cite{CMStch8lhc}. It is also in excellent agreement with the NNLO result in \cite{NNLOtch}. 

We continue with $t$-channel aNNLO $p_T$ distributions \cite{NKtchpt}. In the right plot of Fig. \ref{figsingletop} we display the top $p_T$ distributions in $t$-channel production at various LHC energies.

\begin{figure}
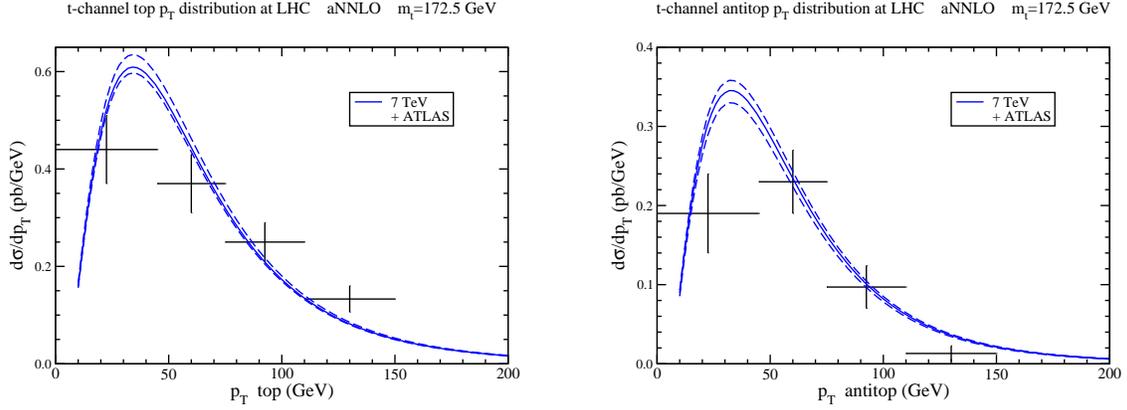

\begin{center}
\includegraphics[width=.45\textwidth]{pttoptchATLASplot.eps}
\hspace{10mm}
\includegraphics[width=.45\textwidth]{ptantitoptchATLASplot.eps}
\caption{Top (left) and antitop (right) $p_T$ distributions in $t$-channel production at 7 TeV LHC energy compared with ATLAS data \cite{ATLAStch7lhc}.}
\label{figpttoptchATLAS}
\end{center}
\end{figure}

In Fig. \ref{figpttoptchATLAS} we display the top and antitop $p_T$ distributions in single-top $t$-channel production at 7 TeV LHC energy and compare with ATLAS data \cite{ATLAStch7lhc}, finding good overall agreement.

\begin{table}[htb]
\begin{center}
\begin{tabular}{c|c|c|c}
LHC  & $t$ &  
${\bar t}$ & Total (pb) \\ 
\hline
8 TeV  & $3.75 \pm 0.07 \pm 0.13$ 
& $1.90 \pm 0.01 \pm 0.08$
& $5.65 \pm 0.08 \pm 0.21$
\\ 
13 TeV & $7.07 \pm 0.13 {}^{+0.24}_{-0.22}$ 
& $4.10 \pm 0.05 {}^{+0.14}_{-0.16}$ 
& $11.17 \pm 0.18 \pm 0.38$
\\ 
14 TeV & $7.79 \pm 0.14 {}^{+0.31}_{-0.24}$ 
& $4.57 \pm 0.05 {}^{+0.18}_{-0.17}$
& $12.35 \pm 0.19 {}^{+0.49}_{-0.41}$
\end{tabular}
\caption{aNNLO single-top and single-antitop $s$-channel cross sections with $m_t=173.3$ GeV.}
\label{tab2}
\end{center}
\end{table}

Next we discuss $s$-channel production at the LHC. In Table \ref{tab2} we show the single-top and single-antitop $s$-channel production cross sections at aNNLO at the LHC, as well as the total sum of the two.
The first uncertainty is again from scale variation while the second is from 
pdf errors with MSTW2008 NNLO pdf at 90\% C.L. \cite{MSTW2008}. 

In the left plot of Fig. \ref{figsingletop} we plot the $s$-channel total cross section as a function of collider energy.
At 8 TeV LHC energy we compare with data from ATLAS \cite{ATLASsch8lhc} and CMS \cite{CMSsch8lhc}. The data from both experiments have much larger uncertainties than the theoretical prediction but are in excellent agreement with it.

\begin{figure}
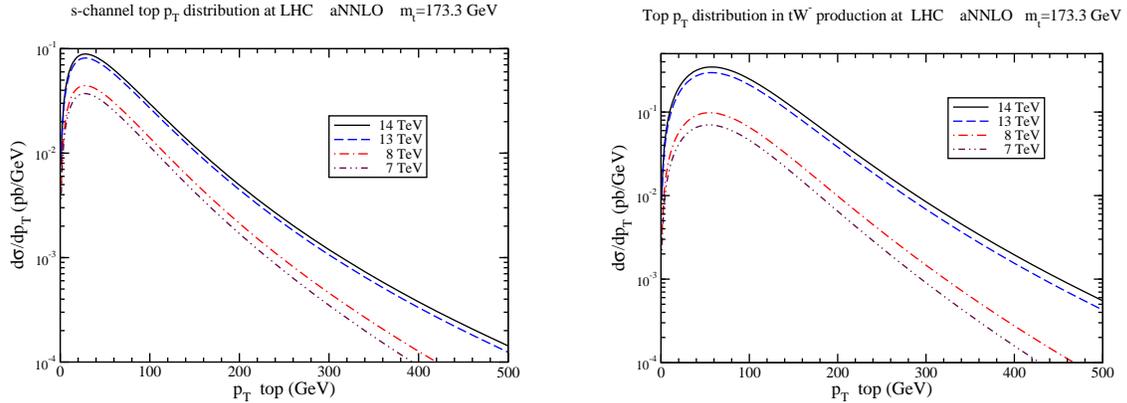

\begin{center}
\includegraphics[width=.45\textwidth]{pttopschlhcplot.eps}
\hspace{10mm}
\includegraphics[width=.45\textwidth]{pttoptWlhcplot.eps}
\caption{Top $p_T$ distributions in single-top $s$-channel production (left) and in $tW^-$ production (right).}
\label{figpttopschtW}
\end{center}
\end{figure}

In the left plot of Fig. \ref{figpttopschtW} we display new aNNLO results for the top $p_T$ distributions in $s$-channel production at LHC energies.

\begin{table}[htb]
\begin{center}
\begin{tabular}{c|c|c}
 LHC  & $tW^-$ &  
$tW^- + {\bar t}W^+$ (pb) \\ 
\hline
 8 TeV  & $11.0 \pm 0.3 \pm 0.7$
& $22.0 \pm 0.6 \pm 1.4$
\\ 
13 TeV & $35.20 \pm 0.9 {}^{+1.6}_{-1.7}$
& $70.40 \pm 1.8 {}^{+3.2}_{-3.4}$
\\ 
 14 TeV & $41.6 \pm 1.0 {}^{+1.5}_{-2.3}$  
& $83.1 \pm 2.0 {}^{+3.1}_{-4.6}$
\end{tabular}
\caption{aNNLO $tW$ cross sections with $m_t=173.3$ GeV.}
\label{tab3}
\end{center}
\end{table}

We continue with associated $tW$ production at aNNLO at the LHC. In Table \ref{tab3} we show the $tW$ production cross sections at aNNLO at the LHC. The first uncertainty is again from scale variation while the second is from MSTW2008 \cite{MSTW2008} pdf errors.

In the left plot of Fig. \ref{figsingletop} we plot the $tW$ total cross section as a function of collider energy.
At 7 TeV LHC energy, we compare with $tW$ data from ATLAS \cite{ATLAStW7lhc} and CMS \cite{CMStW7lhc}.
At 8 TeV LHC energy we compare with ATLAS/CMS combination data \cite{tW8lhccombo}. We find excellent agreement of theory with data for both LHC energies.
In the right plot of Fig. \ref{figpttopschtW} we display the top-quark $p_T$ distributions in $tW^-$ production at LHC energies.

\section{FCNC top production}

Finally, we consider FCNC top-quark production via anomalous gluon couplings.
The partonic processes are of the form $gu \rightarrow tg$ which involve 
$t$-$u$-$g$ couplings. We considered these processes beyond leading order and calculated the soft-gluon corrections at NLL accuracy in \cite{NKEM}. 

The ratio of the LO and aNLO cross sections at various choice scales to the LO result with $\mu=m_t$ was shown in \cite{NKEM} at both 7 and 14 TeV LHC energies. It was found that the NLO soft-gluon corrections are large and they reduce the scale dependence of the cross section. At both energies the NLO soft-gluon corrections increase the LO cross section by around 60\% for $\mu=m_t$. The reduction in scale variation over $m_t/2 \le \mu \le 2m_t$ is also very significant.

\section{Summary}

We have presented results for NNLL soft-gluon corrections for electroweak-boson  and single-top production. The aNNLO corrections are significant 
at the LHC and the Tevatron.
There is excellent agreement for single-top production with LHC and Tevatron data in all channels.
Future work will extend these results to more differential distributions and 
aN$^3$LO.

\end{document}